\RequirePackage{fixltx2e}
\documentclass[aip,jcp,reprint,floatfix]{revtex4-1}

\usepackage{amsmath,amsfonts}
\usepackage{wasysym}
\usepackage{bm}
\usepackage[acronym]{glossaries}
\usepackage{graphicx}
\usepackage{dcolumn}
\usepackage{color}
\usepackage{nicefrac}
\usepackage{ifpdf}
\usepackage{natbib}
\usepackage[byname]{smartref}
\usepackage[breaklinks, %
colorlinks, %
linkcolor=blue, %
citecolor=blue, %
urlcolor=blue]{hyperref}
\usepackage[table]{xcolor}
\usepackage{subfigure}

\DeclareMathOperator{\tr}{\rm Tr}

\newcommand{\meanT}[1]{\ensuremath{\langle#1\rangle}_T}
\newcommand{\ket}[1]{\ensuremath{|#1\rangle}}
\newcommand{\bra}[1]{\ensuremath{\langle #1|}}

\newcommand{\BRA}[1]{\langle{} #1 \vert}
\newcommand{\KET}[1]{\vert{} #1 \rangle}

\newcommand{\eq}[1]{Eq.~(\ref{#1})}
\newcommand{\eqs}[1]{Eqs.~(\ref{#1})}
\newcommand{\fig}[1]{Fig.~\ref{#1}}

\newcommand{\be}{\begin{equation}}
\newcommand{\ee}{\end{equation}}
\newcommand{\bea}{\begin{eqnarray}}
\newcommand{\eea}{\end{eqnarray}}

\newcommand{\HLVC}{H_{\rm LVC}}

\ifpdf %
\fi

\begin{document}

\newacronym{CI}{CI}{conical intersection} %
\newacronym{GP}{GP}{geometric phase} %
\newacronym{LVC}{LVC}{linear vibronic coupling} %
\newacronym{TCLME}{TCLME}{time-convolutionless master equation} %
\newacronym{CNC}{CNC}{collective nuclear coordinates} %
\newacronym{DOF}{DOF}{degrees of freedom} %
\newacronym{PES}{PES}{potential energy surface} %
\newacronym{TDPT}{TDPT}{time-dependent perturbation theory} %

\title{Geometric phase effects in low-energy dynamics near conical
  intersections:  A study of the multidimensional linear vibronic coupling model}
  


\author{Lo{\"i}c Joubert-Doriol} \affiliation{Department of Physical
  and Environmental Sciences, University of Toronto Scarborough,
  Toronto, Ontario, M1C 1A4, Canada; and Chemical Physics Theory
  Group, Department of Chemistry, University of Toronto, Toronto,
  Ontario, M5S 3H6, Canada}

\author{Ilya G. Ryabinkin} %
\affiliation{Department of Physical and Environmental Sciences,
  University of Toronto Scarborough, Toronto, Ontario, M1C 1A4,
  Canada; and Chemical Physics Theory Group, Department of Chemistry,
  University of Toronto, Toronto, Ontario, M5S 3H6, Canada}

\author{Artur F. Izmaylov} %
\affiliation{Department of Physical and Environmental Sciences,
  University of Toronto Scarborough, Toronto, Ontario, M1C 1A4,
  Canada; and Chemical Physics Theory Group, Department of Chemistry,
  University of Toronto, Toronto, Ontario, M5S 3H6, Canada}

\date{\today}

\begin{abstract}
  In molecular systems containing \glspl{CI}, a nontrivial \gls{GP}
  appears in the nuclear and electronic wave-functions in the
  adiabatic representation. We study \gls{GP} effects in nuclear dynamics of
  an $N$-dimensional \gls{LVC} model. The main impact of GP on low-energy nuclear dynamics is reduction of population transfer 
  between the local minima of the LVC lower energy surface. For the LVC model, we proposed an isometric coordinate transformation 
  that confines non-adiabatic effects within a two-dimensional subsystem interacting with an $N-2$ dimensional environment. 
  Since environmental modes do not couple electronic states, all GP effects originate from nuclear dynamics within 
  the subsystem. We explored when the GP affects nuclear dynamics of the isolated subsystem, and how the subsystem-environment interaction 
  can interfere with GP effects.
  Comparing quantum dynamics with and without  \gls{GP} allowed us to devise simple rules to determine 
  significance of the \gls{GP} for nuclear dynamics in this model.  
\end{abstract}


\maketitle

\glsresetall

\section{Introduction}
\label{sec:introduction}

\begin{figure}
  \includegraphics[width=0.5\textwidth]{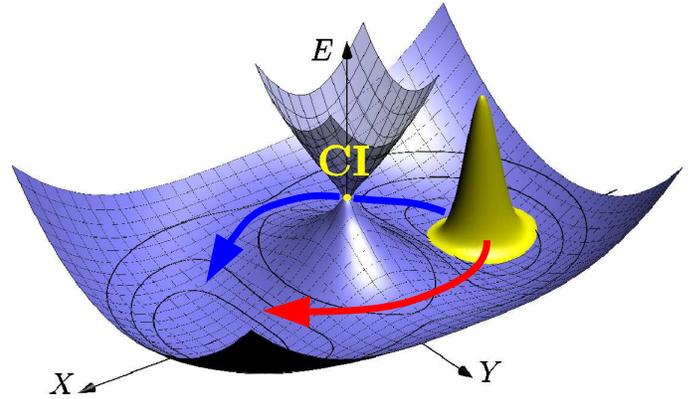}
  \caption{Destructive interference due to geometric phase in low
    energy dynamics. The minimum on the $X<0$ ($X>0$) side can correspond to
    the donor (acceptor) state for charge or energy transfer processes.\cite{Blancafort:2005/JACS/3391,Blancafort:2001/JACS/722}}
  \label{fig:scheme_D-A}
\end{figure}

\Glspl{CI} are known to play a key role in radiationless electronic
transitions in molecular systems \cite{Migani:2004/271}. However, the
electronic transitions are not the only features that \glspl{CI}
introduce in to the nuclear dynamics: Another intriguing, but much less
investigated, aspect of the nuclear dynamics near \gls{CI} is
nontrivial \gls{GP} occurring in adiabatic electronic and nuclear wave-functions on
encircling the \gls{CI} seam~\cite{Berry:1984/rspa/45,
  Mead:1979/jcp/2284}. The \gls{GP} effects can become important even
for nuclear dynamics predominantly confined to a single adiabatic
electronic surface~\cite{Ryabinkin:2013un}.  For example, charge and
energy transfer processes considered in the adiabatic representation
may not go far beyond a single electronic surface description, and thus,
can experience significant \gls{GP} effects
in the presence of \gls{CI}~\cite{Blancafort:2005/JACS/3391,
  Blancafort:2001/JACS/722, Izmaylov:2011/jcp/234106} (see \fig{fig:scheme_D-A}).  As we have
shown in our previous paper~\cite{Ryabinkin:2013un}, the \gls{GP} has
a significant impact on the low-energy nuclear dynamics of systems
with \gls{CI}: tunnelling of a localized nuclear wave-packet from one well to
another is significantly reduced or even blocked completely in the
presence of \gls{GP}. This is a result of destructive interference
between parts of the initial wave packet traveled on different sides
from the \gls{CI} (\fig{fig:scheme_D-A}). The same interference effect
causes a nodal line to appear in the tunnelled wave
packet~\cite{Ferretti:1996/jcp/5517,Schon:1994/cpl/55}.

Most of the previous studies of \gls{GP} were done on small models
\cite{Ryabinkin:2013un,Schon:1995/jcp/9292, Schon:1994/cpl/55,
  Althorpe:2008/jcp/214117} or molecular systems with a few
atoms\cite{Kendrick:1996/jcp/7475, Bouakline:2010/mp/969,
  JuanesMarcos:2005/sci/1227}, therefore, it is still unclear how
\gls{GP} effects can modify dynamics of a large multidimensional
system with \gls{CI}. Usually quantum effects diminish with increase of the system size, and 
\gls{GP} effects as purely quantum are expected to follow this trend. 
Study of \citet{Kelly:2010/jcp/084502} supported this view by
illustrating that the associated with \gls{GP} nodal line in a subsystem adiabatic nuclear density 
disappears after including interaction with an environment. 
On the other hand, studies on two-level spin subsystems
coupled to environment~\cite{Carollo:2003/prl/160402,
  Lombardo:2013/pra/032338} suggest that \gls{GP} effects survive, and
thus, the \gls{GP} can be used to encode information in quantum
computing.  Considering these seemingly controversial results from two
communities, we would like to assess \gls{GP} effects in large
vibronic multidimensional systems with \glspl{CI}.  The difference of
our approach is to consider not only the nodal line in the nuclear
density as a sign of \gls{GP} but also to compare population
dynamics in simulations with and without \gls{GP} effects. The latter
comparison is mostly motivated by the idea that the nodal line is one
of the consequences of \gls{GP} effects but its absence, in general,
cannot be considered as a sign of \gls{GP} insignificance.

To study \gls{GP} effects in multidimensional systems, we
consider the $N$-dimensional \gls{LVC} Hamiltonian
model~\cite{Koppel:1984/acp/59}
\begin{align}
  \label{eq:LVC0}
  \HLVC = & \sum_j^N \frac{1}{2}\left(p_j^2 + \omega_j^2 q_j^2
  \right)\mathbf{1}_2 +
  \begin{pmatrix}
    \kappa_j q_j & c_j      q_j \\
    c_j q_j & \tilde\kappa_j q_j
  \end{pmatrix}\nonumber\\
  &\hspace{1cm}+\begin{pmatrix}
    -\delta/2 & 0 \\
    0 & \delta/2
  \end{pmatrix},
\end{align}
where we use mass-weighted coordinates $q_j$ and their conjugated
momenta $p_j$, $\omega_j$ are the frequencies, $\kappa_j$,
$\tilde\kappa_j$, and $c_j$ are linear coupling constants. $\delta$ is the
energy difference between the minima of the two diabatic electronic potentials.
All quantities in this equation are given in atomic units, which are
used throughout this work. In spite of its simplicity, $\HLVC$ has been
successfully applied to model vibronic spectra of various molecular
systems (e.g., Jahn-Teller distorted
molecules)~\cite{Meng:2013/jcp/014313, Li:2013/jcp/094313,
  Leveque:2013/jcp/044320} and used as an ansatz for approximate
diabatization methods~\cite{Koppel:2001/jcp/2377,
  Allan:2010/jpca/8713, Izmaylov:2011/jcp/234106}.

Another advantage of the \gls{LVC} model found in the course of this
work is existence of an isometric transformation that maps the
$N$-dimensional $\HLVC$ into a Hamiltonian where all non-adiabatic
effects are confined within a two-dimensional branching subspace
spanned by two \gls{CNC}. The other $N-2$ \gls{CNC} can be seen as
environmental \gls{DOF} that interacts with the \gls{CNC} of the
branching subspace identically for both electronic states.  Our
transformation is similar to the ones found earlier by Cederbaum and
coworkers~\cite{Cederbaum:2005/prl/113003, Burghardt:2006/mp/1081,
  Gindensperger:2006/jcp/144103} with the main difference that all
previous transformations had introduced three-dimensional
electronically coupled subsystems while ours has only a
two-dimensional subsystem.  Thus, we employ the methods developed to
analyze \gls{GP} effects in the 2D-LVC problem~\cite{Ryabinkin:2013un}
and augment them by time-convolutionless master equation
approach\cite{Book/Breuer:2002} to account for the
subsystem-environment interaction.  Owing to the isometric transformation that
confines non-adiabatic effects within the branching subspace,
multimode consideration of \gls{GP} effects in $\HLVC$ is split in
two steps: 1) \gls{GP} effects within the branching subspace, and 2)
influence of subsystem-environment interaction on \gls{GP} effects.
This split allows us to formulate simple rules on when \gls{GP}
effects are expected to be important in the $N$-dimensional
\gls{LVC} model.

The rest of the paper is organized as follows. In
Section~\ref{sec:motiv-exampl-two} we illustrate the origin of the
\gls{GP} on a two-dimensional \gls{LVC} model.
Section~\ref{sec:method} describes the \gls{LVC} isometric
transformation and techniques used to simulate the nuclear
non-adiabatic dynamics.  Section~\ref{sec:qual} provides qualitative
analysis of the nuclear non-adiabatic dynamics for several variations
of system parameters.  Section~\ref{sec:res} discusses results of the
nuclear dynamics with and without the \gls{GP} for isolated subsystem
and subsystem interacting with its environment.
Section~\ref{sec:concl} concludes by summarizing our main findings.

\section{Motivating example: two-dimensional LVC model}
\label{sec:motiv-exampl-two}

To illustrate importance of \gls{GP} effects we consider the
simplest two-dimensional \gls{LVC} model where a nontrivial \gls{GP}
appears. The model Hamiltonian is
\begin{equation}
  \label{GPLVC2D} 
  H_{\rm 2D}= T_{\rm 2D}{\mathbf 1}_2
  + \begin{pmatrix}
    V_{11} & V_{12} \\
    V_{12} & V_{22}
  \end{pmatrix},
\end{equation}
where $T_{\rm 2D}=-1/2 (\partial^2 /\partial x^2 +\partial^2 /\partial
y^2) $ is the nuclear kinetic energy operator, $x$ and $y$ are the
mass-weighted coordinates, $V_{11}$ and $V_{22}$ are the diabatic
potentials represented by identical two-dimensional parabolas shifted
in space and coupled by the $V_{12}$ potential
\begin{align}
  \label{eq:diab-me-11}
  V_{11} = {} & \dfrac{\omega^2 }{2}\left[(x+x_0)^2 + y^2\right],
  \quad  V_{12} = c y, \\
  \label{eq:diab-me-22}
  V_{22} = {} & \dfrac{\omega^2}{2} \left[(x-x_0)^2 + y^2\right].
\end{align}
Here, $\omega$ is the frequency for both coordinates, $\pm x_0$ are
the minima of $V_{11}$ and $V_{22}$ potentials, and $c$ is a coupling
constant.  $H_{\rm 2D}$ is an electron-nuclear Hamiltonian written in
the so-called \emph{diabatic representation} with the nuclear kinetic
energy operator $T_{\rm 2D}{\mathbf 1}_2$ diagonal in the electronic
subspace.  Electronic \gls{DOF} in $H_{\rm 2D}$ are abstract vectors
$\KET{1}$ and $\KET{2}$ in a two-dimensional linear space.  Both
non-adiabatic transitions and \gls{GP} effects are accounted in
$H_{\rm 2D}$ implicitly via the off-diagonal elements of the potential
matrix $V_{12}$. To obtain the corresponding \emph{adiabatic
  representation} of the Hamiltonian one needs to diagonalize the
two-state potential matrix in Eq.~(\ref{GPLVC2D}) by unitary
transformation
\begin{equation}
  \label{eq:U-mat}
  U(\theta) = 
  \begin{pmatrix}
    \cos\dfrac{\theta}{2} & -\sin\dfrac{\theta}{2} \\[2ex]
    \sin\dfrac{\theta}{2} & \phantom{-}\cos\dfrac{\theta}{2}
  \end{pmatrix},
\end{equation}
where $\theta$ is a mixing angle between the diabatic states $\KET{1}$
and $\KET{2}$ defined as
\begin{equation}
  \label{eq:theta}
  \theta = \arctan \dfrac{2\,V_{12}}{V_{11} - V_{22}}.
\end{equation}
The diabatic-to-adiabatic transformation $U(\theta)$ defines the adiabatic electronic states 
\begin{eqnarray}
  \KET{\phi_1^\text{adi}} & = & \phantom{-}\cos\frac{\theta}{2}\,\KET{1} +
  \sin\frac{\theta}{2}\,\KET{2} \\ 
  \KET{\phi_2^\text{adi}} & = & -\sin\frac{\theta}{2}\,\KET{1} +
  \cos\frac{\theta}{2}\,\KET{2}
\end{eqnarray}
and brings the
Hamiltonian~(\ref{GPLVC2D}) to the form
\begin{equation}
  \label{eq:adiab}
  {H}_{\rm 2D}^\text{adi} =   
  \begin{pmatrix}
    T_{\rm 2D} + \tau_{11}& \tau_{12} \\
    \tau_{21} & T_{\rm 2D} +\tau_{22}
  \end{pmatrix} +
  \begin{pmatrix}
    W_{1} & 0 \\
    0 & W_{2}
  \end{pmatrix},
\end{equation}
where
\begin{align}
  \label{eq:Wmin}
  W_{1,2} = & {} \dfrac{1}{2}\left(V_{11} + V_{22}\right) \mp
  \dfrac{1}{2}\sqrt{\left(V_{11} - V_{22}\right)^2 + 4 V_{12}^2},
\end{align}
are the adiabatic potentials with the minus (plus) sign for $W_1$ ($W_2$), $\tau_{ij}=-\left\langle
  \phi_i^\text{adi} | \nabla \phi_j^\text{adi}\right\rangle\cdot\nabla
- \frac{1}{2} \left\langle \phi_i^\text{adi} | \nabla^2
  \phi_j^\text{adi}\right\rangle$ are non-adiabatic couplings with
$\nabla = (\partial/\partial x, \partial /\partial y)$.

In the adiabatic representation, if nuclei undergo infinitely slow
(adiabatic) evolution around the \gls{CI} point, $\theta$ changes from
0 to $2\pi$. Since $U(2\pi) = -\mathbf{1}_2$, both adiabatic
electronic states $\{\KET{\phi_i^\text{adi}}\}_{i=1}^2$ change their
signs after encircling the \gls{CI}. This sign change is the result of
acquisition of the nontrivial \gls{GP} during the adiabatic evolution
around a degeneracy
point~\cite{Berry:1984/rspa/45,Mead:1979/jcp/2284}.  This also means
that both adiabatic electronic states
$\{\KET{\phi_i^\text{adi}}\}_{i=1}^2$ are \emph{double-valued}
functions of nuclear coordinates, and that the \gls{CI} point is a
branching point for them.

\begin{figure}
  \centering
  \includegraphics[width=0.5\textwidth]{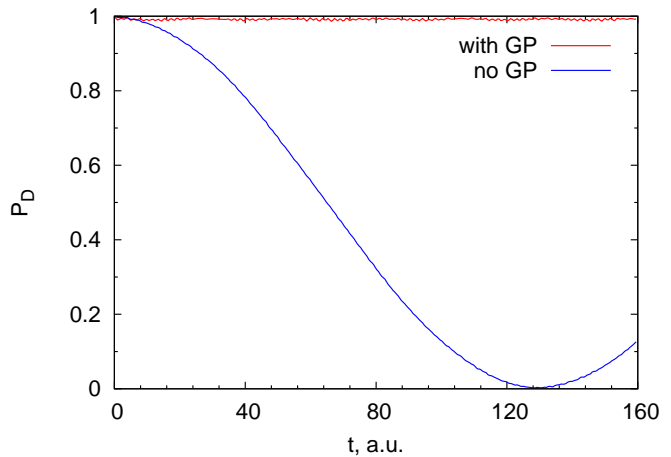}
  \caption{Population transfer between adiabatic wells: $P_D$ is a
    fraction of nuclear density corresponding to negative $x$
    values. Parameters of Eqs.~(\ref{eq:diab-me-11}) and
    (\ref{eq:diab-me-22}) used here are $\omega=2$, $c=3$, $x_0=1.5$
    in a.u. The initial wavefunction is $\KET{\chi}\KET{\phi_1^{\rm
        adi}}$, where $\KET{\chi}$ is the ground vibrational state of
    the Hamiltonian $T_{\rm 2D}+ V_{11}$.}
  \label{fig:dyn}
\end{figure}

\begin{figure}
  \centering
  \includegraphics[width=0.5\textwidth]{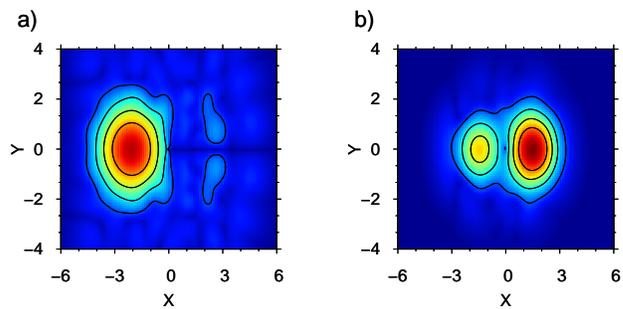}
  \caption{Snapshots of $\sqrt{|\chi^{\rm adi}_1|}$
    at $t = 100.0$ a.u. for the same parameters as in
    Fig.~\ref{fig:dyn}: a) with GP, and b) without
   GP. The square root is used to make a nodal line
    legible.}
  \label{fig:nodes_sym_sub}
\end{figure}

Hamiltonians (\ref{GPLVC2D}) and (\ref{eq:adiab}) should produce
exactly the same nuclear dynamics because they are connected by the
unitary transformation $U(\theta)$.  However, the nuclear
wavefunctions associated with $H_{\rm 2D}$ and $H_{\rm 2D}^\text{adi}$
are subjected to different boundary conditions: In the diabatic
representation both nuclear and electronic wavefunctions are
single-valued. In the adiabatic representation the electronic wave
functions are double-valued, and thus, to have a single-valued
\emph{total} wavefunction, one has to impose the double-valued
boundary condition on the nuclear wave functions $\{\KET{\chi_j^{\rm adi}}\}_{j=1}^{2}$. 
Double-valued boundary conditions can be cumbersome to implement
practically. To address this issue \citet{Mead:1979/jcp/2284} proposed
to factorize the double-valued adiabatic nuclear wavefunctions as
$\KET{\chi_j^\text{adi}} = e^{i\gamma}\,\KET{\tilde{\chi}_j^{\rm adi}}$, where $\gamma$ is a
function that changes from 0 to $\pi$ along a path encircling a
\gls{CI} seam and $\KET{\tilde{\chi}_j^{\rm adi}}$ are single-valued functions. 
The factor $e^{i\gamma}$ is the geometric phase, and for our 2D
model it can be expressed as $\gamma = \theta/2$, since $\theta$ changes by
$2\pi$ upon encircling the CI point. Neglecting the GP or 
double-valuedness of the nuclear wavefunctions in the adiabatic
representation can result in drastically different nuclear dynamics
as shown in Fig.~\ref{fig:dyn}.
Thus, it is \emph{not} sufficient to account only for the
non-adiabatic couplings $\tau_{ij}$ [Eq.~(\ref{eq:adiab})] to obtain
the correct dynamics in the adiabatic representation but the
double-valuedness or \gls{GP} of the nuclear wavefunction must also be
included. The GP $e^{i\gamma}$ makes parts of a nuclear wave packet traveling on 
different sides from the CI point to acquire the opposite phases $e^{\pm i \pi/2}$ 
(red and blue paths in Fig.~\ref{fig:scheme_D-A}). 
This results in the destructive interference between these parts and 
gives rise to a nodal line in the adiabatic nuclear wave packet (Fig.~\ref{fig:nodes_sym_sub}a). 
The GP origin of the nodal line can be verified by simulating nuclear dynamics without 
GP where the nodal line does not appear (Fig.~\ref{fig:nodes_sym_sub}b).

\section{Method}
\label{sec:method}

\subsection{Model Hamiltonian}
\label{sec:n-dimensional-lvc}

To generalize the consideration of the \gls{GP} in the 2D \gls{LVC} model
to the $N$-dimensional case we propose a series of transformations
(detailed in the Appendix A) that bring the $N$-dimensional \gls{LVC}
Hamiltonian to the following equivalent form
\begin{eqnarray}
  \label{eq:LVC2}
  H & = & H_{\rm S} + H_{\rm SB} + H_{\rm B}, 
\end{eqnarray}
with
\begin{eqnarray}
  \label{eq:LVC3a} 
  H_{\rm S\phantom{B}} & = & T_{\rm S}{\mathbf 1}_2 + 
  \begin{pmatrix}
    V_A & V_c \\
    V_c & V_D
  \end{pmatrix},\\
  \label{eq:LVC3b}
  H_{\rm B\phantom{S}} & = & \frac{1}{2} {\sum}_{j=1}^{N-2}
  \left(P_j^2 + \Omega_j^2 Q_j^2\right)\mathbf{1}_2,\\ 
  \label{eq:LVC3c}
  H_{\rm SB} &=& {\sum}_{j=1}^{N-2} (\lambda_{jX} X + \lambda_{jY} Y)
  Q_j\mathbf{1}_2.  
\end{eqnarray}
In this form all non-adiabatic effects are confined in the
two-dimensional subsystem Hamiltonian $H_{\rm S}$ with
\begin{eqnarray}
  \label{eq:def4}
  T_{\rm S} &=&  \frac{1}{2}(P_X^2 + P_Y^2), \\
  \label{eq:def5a}
  V_{D} &=& \frac{1}{2}\left[ \Omega_X^2(X + X_0)^2 +
    \Omega_Y^2(Y + Y_0)^2 + \Delta \right] \\
  \label{eq:def5b}
  V_{A} &=& \frac{1}{2}\left[ \Omega_X^2(X - X_0)^2 +
    \Omega_Y^2(Y - Y_0)^2 - \Delta \right] \\
  \label{eq:def6}
  V_c &=& C_XX+C_YY + \Delta_{12}, 
\end{eqnarray}
where $X$ and $Y$ are the subsystem's \gls{CNC}, $P_X$ and $P_Y$ are
corresponding momenta. The parameters involved in the potentials
$V_{D}$, $V_{A}$, and $V_c$ are functions of the LVC Hamiltonian parameters
and have the following geometrical meaning
\begin{figure}
  \centering
  \includegraphics[width=0.5\textwidth]{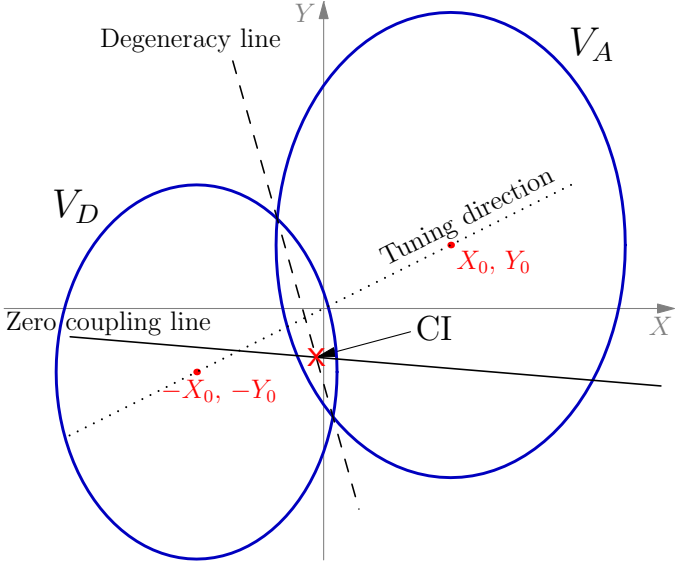}
  \caption{The geometry of the subsystem Hamiltonian~(\ref{eq:LVC3a}):
    the important elements that accompanied the conical intersection
    (terminology is explained in the text) in a constant-energy ($E =
    7$ a.u.) plane. The Hamiltonian parameters are: $\Omega_X = 2$,
    $\Omega_Y = 3/2$, $\mathbf{R}_0 = (3/2, 3/4)$, $\Delta = 3$,
    $\mathbf{C} = (1/4, 3)$, $\Delta_{12} = -7/4$.}
  \label{fig:ham-geometry}
\end{figure}
\begin{itemize}
\item Vector $\mathbf{R}_0 = (X_0,\,Y_0)$ determines a line that
  connects the $V_D$ and $V_A$ minima and is referred as a \emph{tuning
    direction} (Fig.~\ref{fig:ham-geometry}).

\item Vector $\mathbf{G}= (G_X, G_Y) = (\Omega_X^2 X_0, \Omega_Y^2
  Y_0)$ is a normal vector to the \emph{degeneracy line} where $V_D =
  V_A$ (Fig.~\ref{fig:ham-geometry}).
  
\item Vector $\mathbf{C}=(C_X, C_Y)$ determines a \emph{coupling
    direction} and is a normal vector to a \emph{zero coupling line}
  where $V_c = 0$ (Fig.~\ref{fig:ham-geometry}).

\item Parameter $\Delta$ is the energy difference between the $V_D$ and $V_A$
  minima.

\item Parameter $\Delta_{12}$ determines the displacement of the zero
  coupling line along the $\mathbf{C}$ vector.
\end{itemize}
The subsystem coordinates $X$ and $Y$ interact with $N-2$ harmonic
modes $Q_j$ of $H_{\rm B}$ through the diagonal in the electronic
subspace subsystem-bath Hamiltonian $H_{\rm SB}$ with coupling
constants $\lambda_{jX}$ and $\lambda_{jY}$.

For further discussion, we define a special \emph{symmetric} 
subsystem setup that corresponds to the Hamiltonian~(\ref{GPLVC2D}):
$Y_0 = \Delta = \Delta_{12} = C_X = 0$.  
 For this setup both the tuning direction and the zero coupling line coincide with the $X$
direction, whereas the degeneracy line is
orthogonal to the tuning direction and the zero coupling line. 

\subsection{Dynamics of isolated subsystem}
\label{sec:dynam-isol-subsyst}

To assess importance of the GP for the $H_{\rm S}$ Hamiltonian we
model nuclear dynamics in both diabatic and adiabatic representations.
The diabatic Hamiltonian $H_S^{\rm dia}=H_S$ is already given by
Eqs.~(\ref{eq:LVC2})-(\ref{eq:LVC3c}) while the adiabatic Hamiltonian
$H_S^{\rm adi}$ is obtained through diagonalization of the two-state
potential matrix in \eq{eq:LVC3a}.  The propagation of the subsystem
density is done using the unitary evolution $\rho_S(t) =
e^{-iH_St} \rho_S(0) e^{iH_St}$ after projecting $\rho_S(t)$ and $H_S$
on a finite basis set.  The choice of the basis set depends on the
representation of the subsystem Hamiltonian $H_S^{\rm adi}$ or
$H_S^{\rm dia}$ and is explained in the supplemental material.\cite{Doriol:2013un}
All employed basis functions
are single-valued, and thus, effects of the GP are included in the
diabatic representation and are neglected in the adiabatic
representation.  Note that the non-adiabatic transitions between two
electronic states are included in both representations.

To monitor the nuclear dynamics we calculate the time evolution of a
projected subsystem population $P_D(t)=\tr\{\rho_S(t)\hat{P}_D\}$,
where $\hat{P}_D$ is the projector on the donor well which is defined
as $\hat{P}_D=1$ for all $X,Y$ in the left side from the degeneracy
line of \fig{fig:ham-geometry}, and $\hat{P}_D=0$ otherwise.  The
initial state of $\rho_S(t)$ is $\KET{\chi}\KET{\varphi_1^{\rm
    adi}}\BRA{\varphi_1^{\rm adi}}\BRA{\chi}$, where
$\KET{\varphi_1^{\rm adi}}$ is the ground adiabatic state of 
$H_S$, and $\KET{\chi}$ is the ground vibrational state of the
Hamiltonian $T_S+V_D$.

\subsection{Dynamics with environment}
\label{sec:dynam-with-envir}

To account for the interaction $H_{\rm SB}$ between the subsystem and
a large number of bath \gls{DOF} we follow a \gls{TCLME} approach to
reduced subsystem dynamics that accounts for the interaction up to a
second order in $H_{\rm SB}$\cite{Book/Breuer:2002}.  Starting from
the Liouville-von Neumann equation for the total density of the system
\begin{equation}
  {\partial\over{\partial t}}\rho(t){}={}-i\left[H,\rho(t)\right],
\end{equation}
and using standard thermal projectors,\cite{Book/Breuer:2002} one can
integrate out the bath \gls{DOF} assuming that in the zeroth order the
bath density is Boltzmann $\rho_B^T=\exp\left(-\frac{H_{\rm
      B}}{kT}\right)/{\rm Tr}\left\{\exp\left(-\frac{H_{\rm
        B}}{kT}\right)\right\}$, and there is no initial correlation
between the subsystem and environment $\rho(0) = \rho_S(0)\rho_B^T$.
For our subsystem-bath interaction $H_{\rm SB}$, \gls{TCLME} is
\begin{align}
  \label{TCL}
  {\partial\over{\partial t}}\rho_S(t){}=&-i \left[ H_{\rm S} ,
    \rho_S(t) \right] \\ \nonumber &\hspace{-1.5cm}- \sum_j
  \left[\Big(\lambda_{jX}X + \lambda_{jY}Y
    \Big),\Big(\lambda_{jX}X_j(t)+\lambda_{jY}Y_j(t)\Big)\rho_S(t)\right]\\
  \nonumber &\hspace{-1.5cm}- \sum_j
  \left[\rho_S(t)\Big(\lambda_{jX}X_j^\dagger(t)+\lambda_{jY}Y_j^\dagger(t)\Big),\Big(\lambda_{jX}X 
    + \lambda_{jY}Y \Big)\right],
\end{align}
where $X_j(t)$ and $Y_j(t)$ are the $X$ and $Y$ operators dressed by
time-dependent functions 
\bea
X_j(t) &=& \int_0^t dt' e^{-iH_{\rm S} t'} X e^{iH_{\rm S} t'} \meanT{Q_j(0)Q_j(t')}, \\
Y_j(t) &=& \int_0^t dt' e^{-iH_{\rm S} t'} Y e^{iH_{\rm S} t'}
\meanT{Q_j(0)Q_j(t')}.  \eea Here, $\meanT{\hdots}$ is a thermal
average over the bath coordinates with the bath Boltzmann density
$\rho_{B}^T$.  The bath correlators $\meanT{Q_j(0)Q_j(t')}$ can be
evaluated analytically\cite{Book/Breuer:2002}
\begin{align}
  \meanT{Q_j(0)Q_j(t')}=&{1\over{2\Omega_j}}\left(e^{-i\Omega_j
      t'}+{2\cos(\Omega_j t')\over{e^{\frac{\Omega_j}{kT}}-1}}\right).
\end{align}

The propagation of the $\rho_S(t)$ is done numerically after
projecting the density matrix and $H_{\rm S}$ on a finite basis set in
the diabatic or adiabatic representations.  The diabatic-to-adiabatic
transformation does not modify the $H_{\rm B}$ and $H_{\rm SB}$
Hamiltonians, and thus, all parts of \eq{TCL} describing the
subsystem-bath interaction are invariant of the electronic
representation.

The parameters of the harmonic bath are generated by an Ohmic spectral
density\cite{Makri:1999/jpcb/2823} \be J(\Omega) = \pi
\sum_{j=1}^{N-2} \frac{\lambda_j^2}{2\Omega_j}\delta(\Omega-\Omega_j),
\ee where $\lambda_j = \Omega_j\sqrt{\xi\Omega_0}$,
$\Omega_j=-\Omega_c\ln{(1-j\Omega_0/\Omega_c)}$, and
$\Omega_0=\Omega_c(1-e^{\Omega_{\rm max}/\Omega_c})/(N-2)$.  The bath
is characterized by a cut-off frequency $\Omega_c$ that determines the
peak and width of $J(\Omega)$, and a Kondo parameter $\xi$ that
characterizes the overall subsystem--bath coupling strength.  In our
simulations we couple only one subsystem coordinate to the bath modes
therefore one Kondo parameter generating either
$\lambda_{jX}=\lambda_{j}$ or $\lambda_{jY}=\lambda_{j}$ is
sufficient.  As in Ref.~\onlinecite{Kelly:2010/jcp/084502}, we use
$N-2=100$ bath \gls{DOF}, and the highest frequency of the discrete
bath $\Omega_{\rm max} = 3\Omega_c$.  Temperature of the bath is fixed
to 0 K in all simulations to avoid technical complications with the
basis size set in the subsystem dynamics.

\section{Qualitative analysis}
\label{sec:qual}

Before considering the numerical simulations we would like to present
some qualitative analysis of the population transfer dynamics and the
$Y=0$ nodal line emergence for the isolated and coupled to the
environment subsystems.  Our analysis is based on the \gls{TDPT} in
the diabatic representation that takes the population on the donor
diabatic state ($V_D$) as a measure of the population transfer.
Although this measure is formally different from $P_D(t)$ defined
earlier it is qualitatively the same for low-energy dynamics. For
simplicity, we consider pure initial states $\KET{0 0}_D$ for the
isolated subsystem and $\KET{0 0 \mathbf{n}}_D$ for the full
system. Here, we use the state notations $\KET{n_Xn_Y}_{e}$ and
$\KET{n_Xn_Y\mathbf{n}}_{e}$, where $e=D,A$ is an electronic state,
$n_X$ and $n_Y$ are the numbers of vibrational quanta on the $X$ and
$Y$ modes, and $\mathbf{n}$ is a vector of vibrational quanta on the
bath modes $Q_j$.  The subsystem is assumed to have isotropic
parabolas $\Omega_X=\Omega_Y=\Omega$ with $\mathbf{R}_0 = (R/2, 0)$
and $\mathbf{C}\cdot\mathbf{R}_0 = 0$ [Eqs.~(\ref{eq:def5a})
and~(\ref{eq:def5b})].

\subsection{Isolated subsystem}
\label{sec:isolated-subsystem}

The subsystem Hamiltonian~(\ref{eq:LVC3a}) is partitioned as $H_{\rm
  S} = H_0 + V$ with
\begin{eqnarray}
  H_0 & = &  T_{\rm S}{\mathbf 1}_2 + 
  \begin{pmatrix}
    V_A & 0 \\
    0 & V_D
  \end{pmatrix}, \\
  \label{Vf} 
  V &=&   \begin{pmatrix}
    0 & V_c \\
    V_c & 0
  \end{pmatrix}.
\end{eqnarray}
In the first order of \gls{TDPT}, the population transfer for the
symmetric setup considered in Sec.~\ref{sec:n-dimensional-lvc} is \bea\label{P10}
P_{D\rightarrow A}^{(1,a)}(t) &=& \left\vert {}_A\BRA{01} C_YY\sigma_x
  \KET{00}_{D}\right\vert^2 \frac{\sin^2(\Omega t/2)}{(\Omega/2)^2}.
\eea This expression explains both the origin of the nodal line due to
the wave-function parity change along the $Y$ direction and slow
(almost frozen) population transfer in Fig.~\ref{fig:dyn} due to the
damping prefactor $\left\vert {}_A\BRA{01}
  C_YY\sigma_x\KET{00}_{D}\right\vert^2/\Omega^2$.  The population
transfer changes if we depart from the symmetric setup.  Here,
we will consider the following symmetry breaking scenarios: i)
$\Delta\ne0$, ii) $\Delta_{12}\ne0$, and iii) $C_X\ne0$.
\begin{figure}
  \centering
  \includegraphics[width=0.5\linewidth]{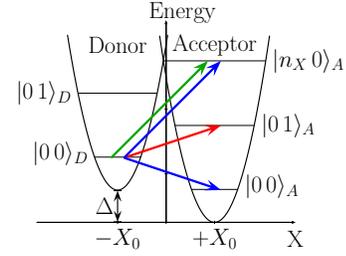}
  \caption{Main channels for population transfer in the diabatic
    representation for a non-symmetric isolated subsystem: i) the red
    arrow for $\Delta\ne0$, ii) the blue arrows for $\Delta_{12}\ne0$,
    iii) the green arrow for $C_X\ne0$.}
  \label{fig:raman-0}
\end{figure}

(i) Energy bias $\Delta\ne 0$ (Fig.~\ref{fig:raman-0}) results in
modification of the population transfer as
\begin{eqnarray}
  \label{P11}
  P_{D\rightarrow A}^{(1,a)}(t) &=& \left\vert {}_A\BRA{01} C_YY\sigma_x
    \KET{00}_{D}\right\vert^2 \\ \notag &\times& 
  \frac{\sin^2[(\Omega-\Delta) t/2]}{[(\Omega-\Delta)/2]^2}.  
\end{eqnarray}
Qualitative difference between the population dynamics in \eqs{P11}
and (\ref{P10}) occurs for the resonance condition $\Delta=\Omega$,
where the population transfer in \eq{P11} can be further simplified as
\begin{equation}
  \lim_{\Delta\to\Omega}P_{D\rightarrow A}^{(1,a)}(t) = \left\vert
    {}_A\BRA{01} C_YY\sigma_x\KET{00}_{D}\right\vert^2 t^2.  
\end{equation}
The resonance strongly facilitates the population flow and corresponds
to the isoenergetic position of $H_0$ vibronic levels that are coupled
by $V_c$.

(ii) Constant coupling $\Delta_{12}\ne 0$ (Fig.~\ref{fig:raman-0})
leads to opening another population transfer channel that in the first
order contributes as
\begin{eqnarray}
  \label{P12} 
  \nonumber  P_{D\rightarrow
    A}^{(1,b)}(t) & = & \Delta_{12}^2 \sum_{n_X=0}^{\infty} |{}_A\BRA{n_X
    0}\sigma_x \KET{00}_{D}|^2 \\ 
  &\times& \frac{\sin^2(n_X\Omega t/2)}{(n_X\Omega/2)^2} .  
\end{eqnarray}
Here, the infinite summation over all vibronic levels with $n_Y=0$ is
due to a finite shift along the $X$ coordinate between the $V_D$ and $V_A$
minima. The resonance condition is satisfied for the $n_X=0$ term
\begin{eqnarray}
  &&\Delta_{12}^2 |{}_A\BRA{n_X 0}\sigma_x \KET{00}_{D}|^2
  \frac{\sin^2(n_X\Omega t/2)}{(n_X\Omega/2)^2}\Bigg{\vert}_{n_X=0} =
  \\ 
  &&\Delta_{12}^2 \left\vert
    {}_A\BRA{01}\sigma_x\KET{00}_{D}\right\vert^2 t^2, 
\end{eqnarray}
therefore, the population transfer is dominated by this term.  Opening
of the $\Delta_{12}\ne 0$ channel provides the population flow that
has a maximum at $Y=0$ line, and thus, its contribution can fill the
nodal line from the channel of \eq{P10}.

(iii) Admixing the tuning direction into $V_c=C_XX+C_YY$ adds a first
order contribution similar to that in \eq{P12}
(Fig.~\ref{fig:raman-0})
\begin{eqnarray}
  \nonumber P_{D\rightarrow A}^{(1,c)}(t)
  &=& \sum_{n_X=1}^{\infty} |{}_A\BRA{n_X 0} C_XX \sigma_x
  \KET{00}_{D}|^2 \\ \label{P13} &\times& \frac{\sin^2(n_X\Omega
    t/2)}{(n_X\Omega/2)^2} .  
\end{eqnarray}
Here, due to the equidistant position of the $V_D$ and $V_A$ minima from the
origin, the matrix element ${}_A\BRA{00} C_XX \sigma_x \KET{00}_{D} =
0$ and $n_X$ runs from 1 rather than from 0 as in \eq{P12}.  Also, due
to this symmetry we expect all integrals ${}_A\BRA{n_X 0} C_XX
\sigma_x\KET{00}_{D}$ with even $n_X$ to contribute insignificantly.
Since $n_X>0$, the resonance condition cannot be satisfied in
\eq{P13}, and the population transfer is similar to that in \eq{P10}.
For the nodal line, this channel has the same effect as the
$\Delta_{12}\ne 0$ channel [\eq{P12}].

\subsection{Interaction with environment}
\label{sec:inter-with-envir}

\begin{figure}
  \centering
  \begin{subfigure}
    \centering
    \includegraphics[width=0.5\linewidth]{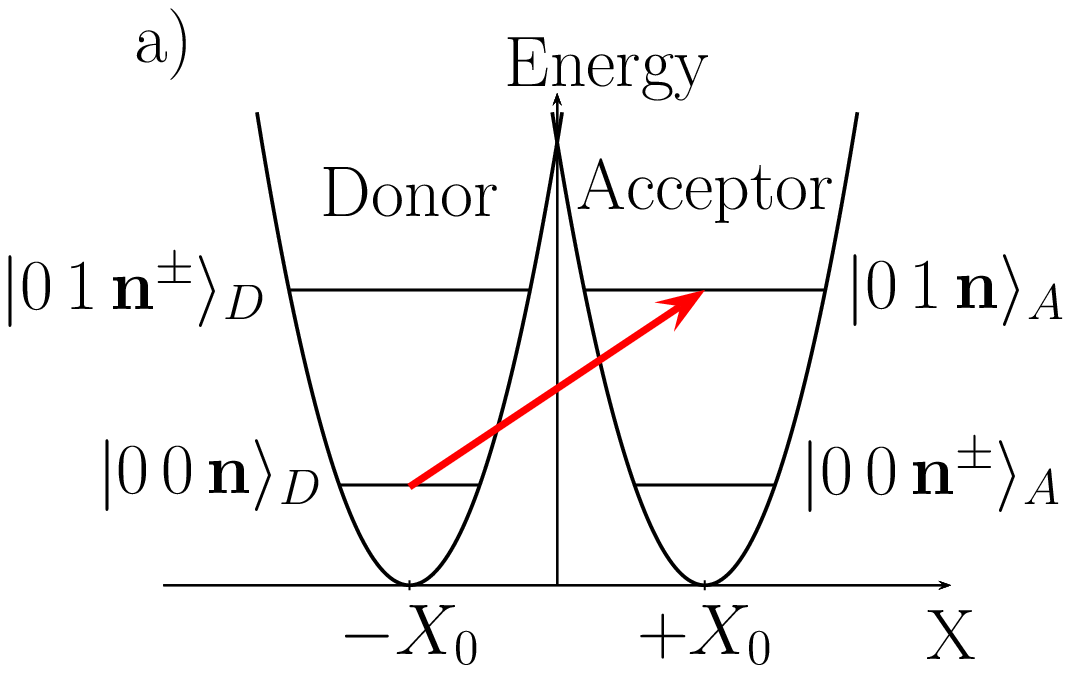}
  \end{subfigure}
  \begin{subfigure}
    \centering
    \includegraphics[width=0.5\linewidth]{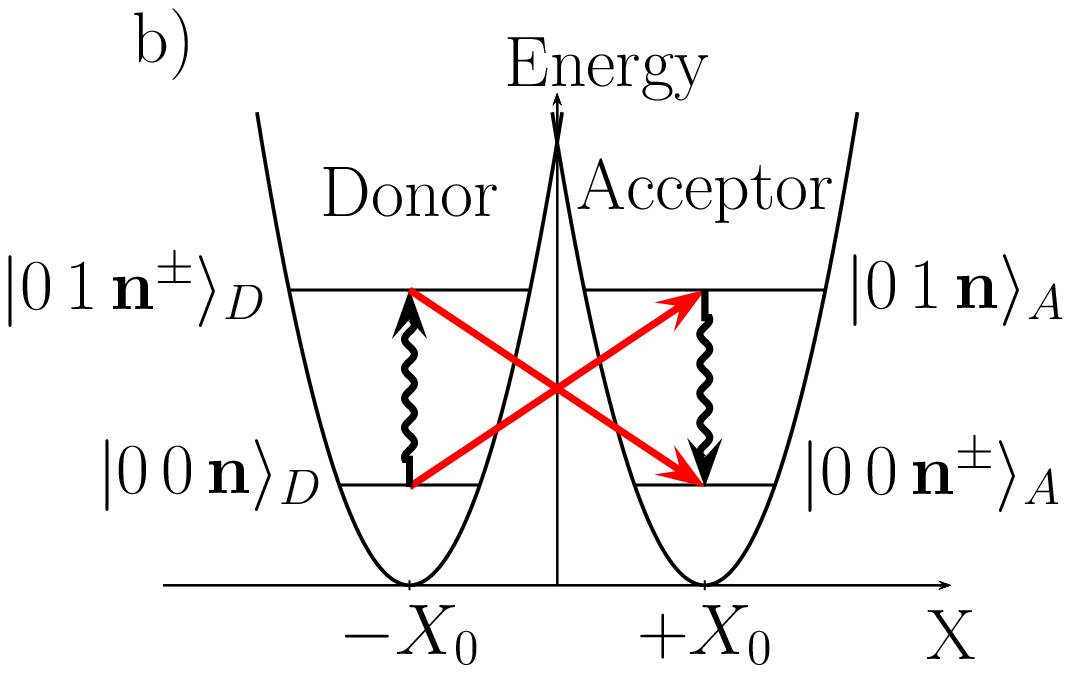}
  \end{subfigure}
  \caption{Main channels for population transfer in the diabatic
    representation for the subsystem interacting with environment: a)
    the first order of TDPT \eq{P1}, b) the second order of TDPT
    \eq{P2}.}
  \label{fig:raman}
\end{figure}

The full Hamiltonian~(\ref{eq:LVC2}) for the symmetric setup can be
written as $H = H_0 + V$ with
\begin{eqnarray}
  \label{Hf}
  H_0 & = & H_{\rm B} + H_{\rm S} - \sigma_x  C_YY, \\ \label{Vf} 
  V &=& \sigma_x C_YY + H_{\rm SB}.
\end{eqnarray}
Adding the environment does not affect the inter-electronic transition
in the first order of \gls{TDPT} (Fig.~\ref{fig:raman}a)
\begin{eqnarray}
  \label{P1} P_{D\rightarrow A}^{(1)}(t) &=& \left\vert
    {}_A\BRA{01\mathbf{n}} C_YY\sigma_x
    \KET{00\mathbf{n}}_{D}\right\vert^2 \frac{\sin^2(\Omega
    t/2)}{(\Omega/2)^2}.  
\end{eqnarray}
The effect of $H_{\rm SB}$ on the donor-acceptor transition appears
only in the second order
\begin{eqnarray}
  \notag P_{D\rightarrow A}^{(2,\pm)}(t)
  &=&\Big{\vert}\int_0^{t}d\tau \int_0^{\tau}d\tau'
  \Big{\{}{}_A\BRA{00\mathbf{n^\pm}} C_YY\sigma_x
  \KET{01\mathbf{n^\pm}}_{D}\\ \notag &\times&
  {}_D\BRA{01\mathbf{n^\pm}} \lambda_{jY}YQ_j
  \mathbf{1}_2\KET{00\mathbf{n}}_{D}
  e^{-i\Omega\tau-i(\Omega\pm\Omega_j)\tau'} \\\notag &+&
  {}_A\BRA{00\mathbf{n^\pm}} \lambda_{jY}
  YQ_j\mathbf{1}_2\KET{01\mathbf{n}}_{A}
  e^{-i\Omega\tau'-i(\Omega\pm\Omega_j)\tau}\\\label{P2} &\times&
  {}_A\BRA{01\mathbf{n}}C_YY\sigma_x
  \KET{00\mathbf{n}}_{D}\Big{\}}\Big{\vert}^2, 
\end{eqnarray}
where $\mathbf{n^\pm}$ and $\mathbf{n}$ differ only by the number of
vibrational quanta along the $Q_j$ mode: $n_j^{\pm} = n_j \pm 1$.
Figure \ref{fig:raman}b illustrates the two components of the integrand
sum in \eq{P2} as two pathways involving energy transfer between the
subsystem and environemnt.  These pathways do not require altering the
parity of the nuclear wave-function along the $Y$ coordinate, and
thus, their contributions do not form the $Y=0$ nodal line.
Therefore, if the transfer due to \eq{P2} is significant compare to
that due to \eq{P1}, the $Y=0$ nodal line in the subsystem wave-packet
will disappear and the donor-acceptor population transfer will be
enhanced. 
Presence of an extra oscillating exponential factor in \eq{P2}
indicates that to have an efficient population transfer along this
channel the bath frequency $\Omega_j$ should be close to the coupling
coordinate frequency $\Omega$.

The difference between pathways with and without bath involvement is
very similar to that between one-photon absorption and Raman
scattering: here, an electronic donor-acceptor transition can be seen as
an absorption of a photon, and energy exchange with the bath is analogous 
to photon scattering.  Also, due to the scattering mechanism the normal dipole
selection rules that require changing the parity of the nuclear
wave-function are altered in the Raman process.

The subsystem-bath interaction through $X$ coordinate ($\lambda_{jX}
XQ_j\mathbf{1}_2$) does not appear in the lowest orders of TDPT for the donor-acceptor transfer probabilities, and
therefore, these terms do not appreciably change the nuclear dynamics
in the diabatic representation.

\section{Results and Discussion}
\label{sec:res}

\subsection{Isolated subsystem}
\label{sec:subsys}

\gls{GP} effects are most prominent in the setup where destructively
interfering parts of the nuclear wave-packet have equal
amplitudes~\cite{Ryabinkin:2013un}. This is the case for
the symmetric setup defined in Sec.~\ref{sec:n-dimensional-lvc}. In what follows we address the
question whether \gls{GP} effects survive if we break the symmetry
between the transfer paths by altering the parameters of the subsystem
Hamiltonian~(\ref{eq:LVC3a}). Staying within the isotropic $\Omega_X =
\Omega_Y=\Omega$ case 
  there are two
scenarios of symmetry breaking with non-symmetric tunnelling paths: i)
making $\mathbf{C}$ and $\mathbf{G}$ non-orthogonal :
$\mathbf{C}\cdot\mathbf{G} \ne 0$ [case (iii) in
Sec.~\ref{sec:isolated-subsystem}], ii) setting $\Delta_{12} \ne
0$ [case (ii) in Sec.~\ref{sec:isolated-subsystem}].  We do not
consider $\Delta$ variations because they do not cause symmetry
breaking between the transfer paths, also their effect has been
studied previously.\cite{Ryabinkin:2013un} All other Hamiltonian
parameters have been assigned the following values: $\Omega = 2$,
$\mathbf{R}_0 = (3/2, 0)$, $\Delta = 0$.

\begin{figure}
  \centering
  \includegraphics[width=0.5\textwidth]{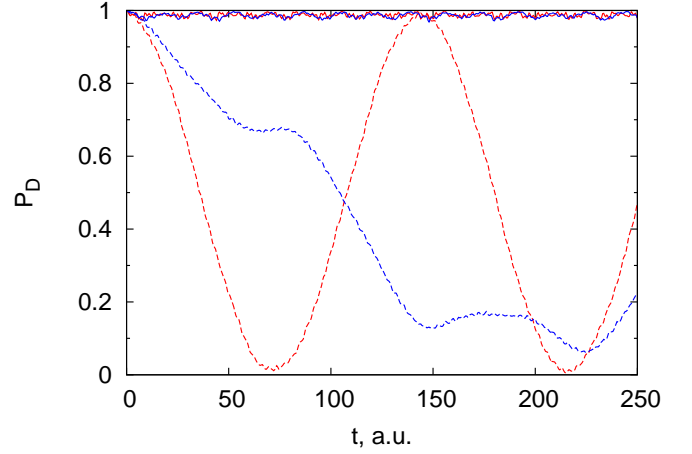}
  \caption{The subsystem donor well population dynamics $P_D(t)$ for different values of $C_X$
    and $C_Y = 4$: (solid red) $C_X=0$ with GP, (solid blue) $C_X=2$ with GP, 
    (dashed red) $C_X=0$ without GP, (dashed blue) $C_X=2$ without GP.} 
  \label{fig:case1}
\end{figure}

{\it(i) $\mathbf{C}\cdot\mathbf{G} \ne 0 : $}
\label{sec:infl-electr-energy}
Starting from the symmetric setup we change the angle between the 
vectors $\mathbf{C}$ and $\mathbf{G} $ by
increasing the component $C_X$ of the vector $\mathbf{C}$ while
keeping $C_Y$ constant. According to our qualitative
analysis, Sec.~\ref{sec:isolated-subsystem}, $C_X$ does not
appreciably affect the dynamics in the diabatic representation.
However, in the adiabatic representation, $C_X$ component deepens the
potential wells resulting in the suppression of the population
transfer in simulations without GP (Fig.~\ref{fig:case1}).
Thus, although the non-orthogonality between $\mathbf{C}$ and
$\mathbf{G}$ preserves the difference between population dynamics with
and without \gls{GP}, it reduces that difference for initial times.

\begin{figure}
  \centering
  \includegraphics[width=0.5\textwidth]{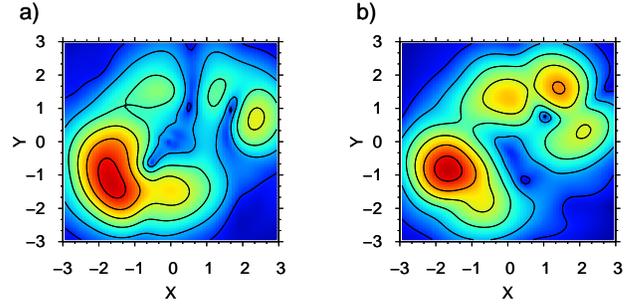}
  \caption{Snapshots of ${\bra{\phi_1^\text{adi}}\rho_S(t)\ket{\phi_1^\text{adi}}}^{1/4}$
    at $t = 15.0$ a.u. for a non-symmetric configuration $C_X = 2$, $C_Y = 6$: a)
    with GP, b) without GP.}
  \label{fig:node-roaming}
\end{figure}

In the symmetric setup with \gls{GP} there is clear nodal structure of
the part of a density tunneled through the CI
(Fig.~\ref{fig:nodes_sym_sub}a). To clearly observe the nodal pattern
in simulations with non-zero $C_X$, we also increased $C_Y$ to facilitate the transfer. As follows from
Fig.~\ref{fig:node-roaming} and our qualitative analysis increasing
$C_X$ destroys the node almost completely. A distinct nodal line
``dissolves'' in a seemingly chaotic interference pattern of a
time-dependent density. At the same time, there are still differences
in population dynamics between models with and without \gls{GP} as
$C_X$ increases. Thus, the nodal line, which is frequently considered
as manifestation of \gls{GP}, is \emph{not} always a reliable
indicator of \gls{GP} significance.

\begin{figure}
  \centering
  \includegraphics[width=0.5\textwidth]{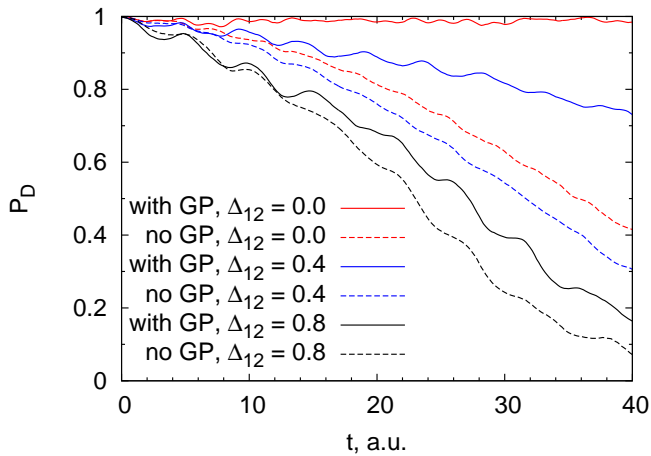}
  \caption{The subsystem donor well population dynamics $P_D(t)$ between equivalent wells 
  for several values of $\Delta_{12}$.}
  \label{fig:case2}
\end{figure}

{\it (ii) $\Delta_{12}\ne 0 : $} The non-zero off-diagonal
coupling constant $\Delta_{12}$ shifts the zero coupling
line. Following the zero coupling line, the \gls{CI} point moves out
of the line connecting two potential minima for $\Delta_{12} \ne
0$. This opens another transfer channel [\eq{P12}] that facilitates
the transfer. For this new channel GP effects are irrelevant and
thus, when this channel becomes dominant the population dynamics with
and without \gls{GP} become similar. To illustrate this idea we gradually increase the value of
$\Delta_{12}$ from 0 to 0.8 a.u. Simulations show
(Fig.~\ref{fig:case2}) that \gls{GP} effects are reduced with
increasing of $\Delta_{12}$.

\begin{figure}
  \centering
  \includegraphics[width=0.5\textwidth]{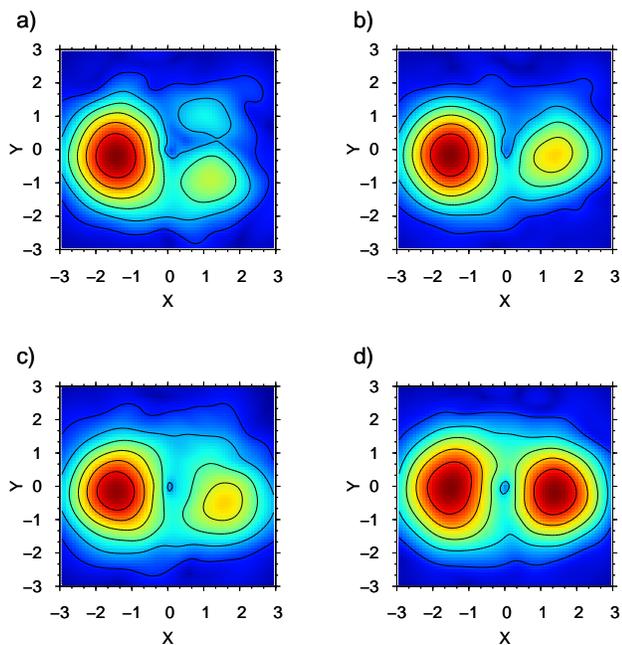}
  \caption{Snapshots of ${\bra{\phi_1^\text{adi}}\rho_S(t)\ket{\phi_1^\text{adi}}}^{1/4}$
    for a non-symmetric configuration $C = 6$, $\Delta_{12} =
    0.6$: a) $t=15$ a.u. with \gls{GP}, b) $t=15$ a.u. without \gls{GP},
    c) $t=30$ a.u. with \gls{GP}, and d) $t=30$ a.u. without \gls{GP}.}
  \label{fig:nodes_d12}
\end{figure}

The nodal line in this set up may form temporarily for the initial
\gls{GP} dynamics, but does not appear later
(Fig.~\ref{fig:nodes_d12}). The reason is that the non-nodal channel
[\eq{P12}] provides the populations transfer to fill the node. In
this case the disappearance of the node correlates well with the
reduction of \gls{GP} significance in the population dynamics.

\subsection{Interaction with environment}
\label{sec:bath}

To observe environmental effects in the most clear settings we
consider the symmetric setup for the subsystem with $\Omega=2$, $C=3$, and $X_0=1.5$ in a.u.,
and two environment-subsystem interaction scenarios: bath modes
are coupled to either the $X$ or the $Y$ coordinate.

\begin{figure}
  \includegraphics[width=0.5\textwidth]{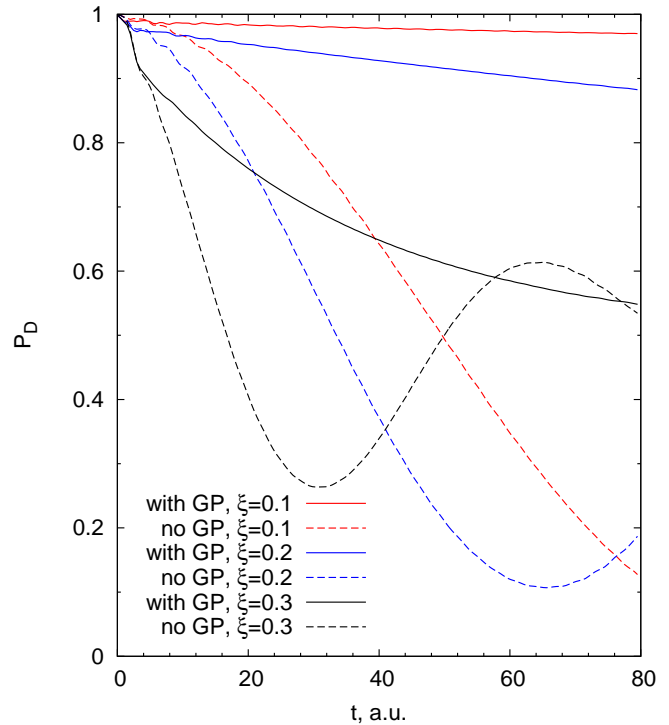}
  \caption{The subsystem donor well population dynamics $P_D(t)$ for $\lambda_{jY}\ne0$ and $\Omega_c=3.5$.}
  \label{fig:bath-Y-T0}
\end{figure}

\textit{(i) Bath is coupled to $Y$
  ($\lambda_{jX}=0$):} Figure~\ref{fig:bath-Y-T0} illustrates the
importance of the GP effects for the donor well population $P_D$ at
various subsystem-bath coupling strengths set with the Kondo parameter
$\xi$.  In both simulation schemes with and without GP, the initial
rate of the population transfer increases with the coupling strength.
In the diabatic simulations this is a result of the new population
transfer pathway [see \eq{P2} and \fig{fig:raman}b] that is opened
because of the subsystem-bath interaction.  In the adiabatic
representation, introducing the subsystem-bath couplings can be seen
as a modification of the subsystem Hamiltonian that involves random
fluctuations changing the transition barrier heights on the lower
adiabatic surface. These barrier fluctuations increase the initial
transfer rate.  On the other hand, increasing the subsystem-bath
coupling strength decreases the amplitude of the population transfer
in the adiabatic representation. We attribute this effect to faster
decoherence that spreads the subsystem wave packet under the influence
of the environment.

\begin{figure}
  \centering
  \includegraphics[width=0.5\textwidth]{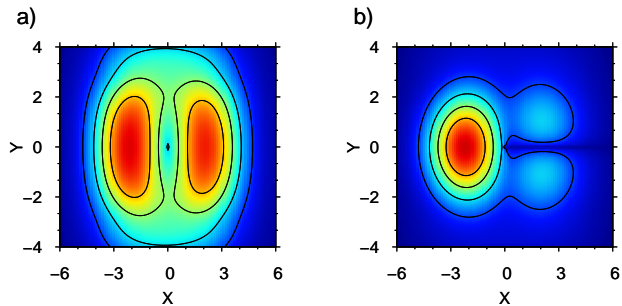}
  \caption{Snapshots of ${\bra{\phi_1^\text{adi}}\rho_S(t)\ket{\phi_1^\text{adi}}}^{1/4}$
    at $t=79.5$ a.u., $\Omega_c=3.5$: a) $\xi=0.3$ with GP and
    $\lambda_{jY}\ne 0$, b) $\xi=0.015$ with GP and $\lambda_{jX}\ne
    0$.}
  \label{fig:node_2Dbath}
\end{figure}

According to our qualitative analysis [\eq{P2}, \fig{fig:raman}b], the
$Y=0$ nodal line disappears when the bath is coupled to the subsystem
coordinate $Y$. A snapshot of the subsystem density given in Fig.~\ref{fig:node_2Dbath}a confirms the node
disappearance, which also agrees with results of Ref.~\onlinecite{Kelly:2010/jcp/084502}.

\begin{figure}
  \includegraphics[width=0.5\textwidth]{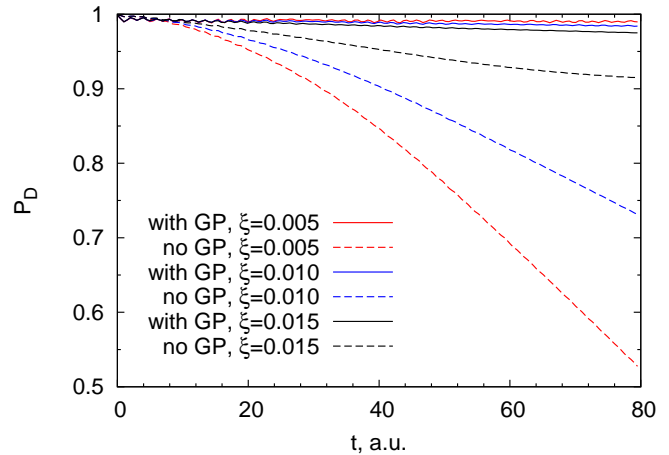}
  \caption{The subsystem donor well population dynamics $P_D(t)$ for small values of
    the Kondo parameter with $\lambda_{jX}\ne0$ and $\Omega_c=3.5$.}
  \label{fig:bath-X-T0}
\end{figure}

\textit{(ii) Bath is coupled to $X$
  ($\lambda_{jY}=0$):} Based on our qualitative analysis, adding a
bath coupled to the $X$ mode does not affect the population transfer
in the diabatic representation. Simulation results in
Fig.~\ref{fig:bath-X-T0} confirms this conclusion: the donor well
population stays almost one as it is in the case of the isolated
subsystem (compare propagations in the diabatic representation in
Figs.~\ref{fig:dyn} and \ref{fig:bath-X-T0}). However, this analysis
cannot be extended to the nuclear dynamics in the adiabatic
representation where increasing the subsystem-bath interaction along
the $X$ mode reduces the population transfer.  This is in line with
the results of one-dimensional tunnelling models in dissipative
environment, where using the instanton approach one can show that a
stronger subsystem-environment coupling reduces the
transfer.\cite{Book/Altland:2006} Although the subsystem-bath coupling
along the $X$ coordinate tends to decrease the differences between
diabatic and adiabatic population transfers, the presence of the nodal
line clearly separates dynamics with and without GP (see
\fig{fig:node_2Dbath}b).

\section{Conclusions}
\label{sec:concl}

We investigated the GP effects in the $N$-dimensional \gls{LVC} model
by reformulating the problem in the subsystem-bath form
Eqs.~(\ref{eq:LVC2}--\ref{eq:LVC3c}). The transformed equations have
all non-adiabatic effects confined within the two-dimensional
subsystem (branching subspace) spanned by tuning and coupling
collective coordinates. The rest $N-2$ collective coordinates form the
harmonic bath, which is bi-linearly coupled with the subsystem
coordinates. After the transformation, the multidimensional character
of the LVC model results in a particular configuration of the
subsystem and a subsystem-bath coupling scheme.  Therefore, impact
of GP in the multidimensional case has been analyzed in two
steps: 1) for the isolated subsystem, and 2) for the subsystem
interacting with its environment.

For the subsystem dynamics, \gls{GP} effects are the most pronounced in
the symmetric configuration: when the tuning and coupling directions
are mutually orthogonal and two electronic state minima do not have
energy difference. For this configuration, the nodal line in the adiabatic density distribution appearing 
in the course of nuclear dynamics is usually considered as the main GP signature. 
All subsystem configurations that break symmetry equivalence of the two population transfer
pathways do not produce the nodal line. However, we found that even in symmetry broken configurations 
GP creates substantial difference in the population dynamics. 
Only if symmetry is broken by increasing the constant inter-electronic coupling $\Delta_{12}$
the GP influence can be reduced.

For the symmetric subsystem configuration we have considered two
subsystem-bath coupling schemes: all modes of the bath are bi-linearly
coupled with either the tuning or the coupling coordinate of the
subsystem.
The population dynamics with and without GP are quite different for
both subsystem-bath coupling schemes.  Therefore, we conclude that the
\gls{GP} effects can survive in a large multidimensional molecular
system. The main difference between the two coupling schemes is
that the nodal line in the subsystem density 
disappears after including the coupling to the bath along the coupling direction
and is preserved in the coupling scheme involving the tuning
direction.  

To summarize, we would like to emphasize that the loss of the
nodal line in subsystem nuclear dynamics does not necessarily mean
insignificance of \gls{GP} effects, and that the most
straightforward and accurate way to assess the \gls{GP} impact is to
compare the time evolution of the quantity of interest in dynamics
with and without \gls{GP}.

\section{Acknowledgments}
We are grateful to R. Kapral and C.-Y. Hsieh for stimulating
discussions.  This work was supported by Natural Sciences and
Engineering Research Council of Canada through the Discovery Grants
Program, and the European Union Seventh Framework Programme
(FP7/2007-2013) under grant agreement PIOF-GA-2012-332233.


\appendix

\section{The effective modes construction}
\label{app:eff-trans}

Here we detail steps of the Hamiltonian transformation starting from Eq.~(\ref{eq:LVC0}) and leading to Eq.~(\ref{eq:LVC2}).

\paragraph{Coordinate translation.---}

We apply a coordinate translation: $q_j=x_j-\frac{\kappa_j+\tilde\kappa_j}{2\omega^2_i}$ to the \gls{LVC} Hamiltonian, Eq.~(\ref{eq:LVC0}), and obtain
\begin{align}
  \label{eq:ALVC1}
  H_1 = & \sum_j^N \frac{1}{2}\left(p_j^2 + \omega_j^2 x_j^2 \right)\mathbf{1}_2 +
  \begin{pmatrix}
   -d_j x_j & c_j x_j \\
    c_j x_j & d_j x_j
  \end{pmatrix}\nonumber \\
&\hspace{1cm}+\begin{pmatrix}
   -\Delta/2 &  \Delta_{12} \\
     \Delta_{12} & \Delta/2
  \end{pmatrix},
\end{align}
where
\begin{eqnarray}
\Delta&=&\sum_j^N\frac{\kappa_j^2-\tilde\kappa_j^2}{2\omega^2_j}+\delta,\nonumber\\
\Delta_{12}&=&\sum_j^N c_j\frac{\kappa_j+\tilde\kappa_j}{2\omega^2_j},\nonumber\\
d_j&=&\frac{\tilde\kappa_j-\kappa_j}{2}.\nonumber
\end{eqnarray}

\paragraph{Subsystem-bath separation.---}

We define a new set of coordinates $\{\tilde x_1,\tilde x_2,\hdots\}$ obtained from $\{x_1,x_2,\hdots\}$ by an orthogonal transformation $\mathbf{O}_1$:
$\mathbf{\tilde x}=\mathbf{O}_1\mathbf{x}$, we use hereafter bold letters as the vector of their corresponding indexed quantities.
$\tilde x_1$ and $\tilde x_2$ are the subsystem coordinates given by
\begin{align}
  \tilde x_1 {} =& \mathbf{e}_d\cdot\mathbf{x}, \nonumber\\
  \tilde x_2 {} =& (\mathbf{c}\cdot\mathbf{x}-\tilde c_1 \mathbf{e}_d\cdot\mathbf{x})/\tilde c_2,
\end{align}
where
\begin{align}
  \mathbf{e}_d {} =& \mathbf{d}/||\mathbf{d}||, \nonumber\\
    \tilde c_1 {} =& \mathbf{c}\cdot\mathbf{e}_d, \nonumber\\
    \tilde c_2 {} =& \sqrt{||\mathbf{c}||^2 - (\mathbf{c}\cdot\mathbf{e}_d)^2}.
\end{align}
Thus, the two first rows of $\mathbf O_1$ are
\begin{equation}
  \begin{pmatrix}
    \mathbf{e}^T_d\\
    (\mathbf{c}^T-\tilde c_1 \mathbf{e}^T_d)/\tilde c_2
  \end{pmatrix}.
\end{equation}
We also define $\tilde d_1 = ||\mathbf{d}||$ for the later use.
The other coordinates, $\{\tilde x_j,\,j=3,\,\hdots,\,N\}$, are the bath coordinates, and are defined by the other rows of $\mathbf{O}_1$. 
We employ the Gram-Schmidt  orthogonalization procedure to obtain them. 
New coordinates $\{\tilde x_1,\tilde x_2,\hdots\}$ allows us to write the Hamiltonian as
\begin{align}
  \label{eq:ALVC2}
  H_2 = & \frac{1}{2}\left( {\sum}_{j=1}^{N} \tilde p_j^2 + \mathbf{\tilde x}^T\mathbf{\Lambda}\mathbf{\tilde x}\right)\mathbf{1}_2
+\begin{pmatrix}
   - \Delta/2 & \Delta_{12} \\
     \Delta_{12} & \Delta/2
  \end{pmatrix} \nonumber\\
&+\begin{pmatrix}
   -\tilde d_1\tilde x_1 & \tilde c_1\tilde x_1 + \tilde c_2\tilde x_2 \\
    \tilde c_1\tilde x_1 + \tilde c_2\tilde x_2 & \tilde d_1\tilde x_1
  \end{pmatrix},
\end{align}
where $\mathbf{\Lambda}$ is the Hessian matrix of both diabatic states. In general,  $\mathbf{\Lambda}$ is non-diagonal in both subsystem and bath subspaces.

\paragraph{Diagonalization of subsystem and bath Hessians.---}

We diagonalize the blocks of the Hessian matrix $\mathbf{\Lambda}$  corresponding to the subsystem and the bath coordinates. This transformation leads to new 
subsystem $\{X,\,Y\}$ and bath $\{Q_j,\,j=1,\,2,\,\hdots,\,N-2\}$ coordinates. The total Hamiltonian can be split into three parts 
\begin{align}
  \label{eq:ALVC3} H_3 = H_{\rm S} + H_{\rm SB} + H_{\rm B},
\end{align}
where
\begin{eqnarray}
H_{\rm S\phantom{B}} &=& {1\over2}\left( P_X^2 +  \Omega_X^2 X^2 + P_Y^2 +  \Omega_Y^2 Y^2\right)\mathbf{1}_2  \nonumber\\
&&\hspace{-0.75cm}+  \begin{pmatrix}
   - \Delta/2 &  \Delta_{12} \\
     \Delta_{12} &  \Delta/2
  \end{pmatrix}
+ \begin{pmatrix}
   -G_X X - G_Y Y & C_X X + C_Y Y \\
    C_X X + C_Y Y & G_X X + G_Y Y    
  \end{pmatrix},\nonumber\\
H_{\rm B\phantom{S}} &=& \frac{1}{2} {\sum}_{j=1}^{N-2} \left(P_j^2 +  \Omega_j^2 Q_j^2\right)\mathbf{1}_2, \nonumber\\
H_{\rm SB} &=& {\sum}_{j=1}^{N-2}  ( \lambda_{jX} X +  \lambda_{jY} Y) Q_j\mathbf{1}_2.
\end{eqnarray}
Here, all non-adiabatic couplings are confined in the subsystem Hamiltonian $H_{\rm S}$ and coefficients $G_{X/Y}$ and $C_{X/Y}$ are obtained from $\tilde d_1, \tilde c_1, \tilde c_2$ [\eq{eq:ALVC2}] by the orthogonal transformation of the subspace coordinate. The environment part $H_{\rm B}$ constitutes a harmonic bath that interacts with the subsystem $H_{\rm S}$
by $H_{\rm SB}$.  $H_{\rm SB}$ has simple bilinear terms with coupling constants $\lambda_{jX}$ and $\lambda_{jY}$ that are independent of electronic state.
$H_{\rm S}$ can be further simplified by completing the squares in $X$ and $Y$ coordinates, 
introducing the minima $X_0=G_X/\Omega_X^2$ and $Y_0=G_Y/\Omega_Y^2$, and neglecting a global energy shift
\begin{align}
H_{\rm S} &= T_{\rm S}{\mathbf 1}_2
+  \begin{pmatrix}
   V_A & V_c \\
   V_c & V_D
  \end{pmatrix},
\end{align}
where $T_{\rm S}$, $V_A$, $V_D$, and $V_c$ are defined by Eqs.~(\ref{eq:def4}-\ref{eq:def6}) given in the main text.

%

\end{document}